\documentclass[11pt]{article}
\usepackage{fixltx2e}
\usepackage{hyperref}
\pdfoutput=1

\begin{document}

\title{Ballistic Impact of Dense Particle Suspensions}

\author{Bradley J. Marr, Oren E. Petel, Andrew J. Higgins,\\ and David L. Frost
\\ McGill University, Department of Mechanical Engineering\\ Montr\'{e}al, Qu\'{e}bec, Canada \and Simon Ouellet
\\ Defence Research and Development Canada - Valcartier\\ Val-B\'{e}lair, Qu\'{e}bec, Canada}

\maketitle

\begin{abstract}
The ballistic impact of various dense particle suspensions is of interest for the development of superior materials for personal protective equipment.
The dynamic response of the fluids under impact of a fragment simulating projectile at various incident velocities was examined for this purpose. High-speed fluid dynamic videos of these ballistic impacts were used to analyze the effects of various suspension parameters on the response of the fluids. It was found experimentally that the shear thickening behaviour of the suspensions dominated the response at low incident velocities, but the results converge based on density at higher impact velocities.
\end{abstract}

\section*{Introduction}

Dense particle suspensions consist of solid particles suspended in a carrier fluid. Many of these dense particles suspensions have a shear thickening characteristic under dynamic loading and are classified as Shear Thickening Fluids (STFs). STFs are non-Newtonian fluids whose viscosity increases under increasing shear rates. Under low shear rates, a STF provides little resistance to motion, whereas at higher shear rates the fluid exhibits an increase in viscosity, some fluids can undergo a discontinuous response with an infinite effective viscosity, similar to a solid. This field-responsive behaviour of STFs is advantageous for protective equipment$^{1,2}$, whereby the fluid can behave differently under various strains. Under the strains associated with regular motions, the fluid is able to flow and the armour would be flexible. In the event of a high strain rate event, the armour would stiffen and provide protection. Integration of STFs embedded in traditional ballistic fabrics is proposed for the development of lighter, less constraining, and more effective body armour$^{1,2}$. Similar body armour systems have shown promise in increasing the protective characteristics of the ballistic fabrics against stabbings$^{3}$.

\section*{Experimental Details}
The dynamic response of dense particle suspensions was examined experimentally by measuring the velocity decrement of a NATO standard 17 grain Fragment Simulating Projectile that was fired through a cylindrical specimen of the suspensions. The cylindrical test section, containing the dense particle suspension, used mylar disks to contain the test fluid allowing easy penetration of the projectile; these mylar disks were shown to not influence the experimental results. The velocity of the projectile was measured on entrance and exit of the test sample using a high-speed camera at 13,000 frames per second. The submitted video  is a compilation of the experimental results.

Ethylene glycol was used as the carrier fluid in all of the suspensions. The suspended particles were varied between SiO\textsubscript{2}, known to be readily shear thickening at large volume fractions$^{1,2}$, and SiC, a high density, high strength ceramic. The ballistic response of these dense particle suspensions were compared based on their rheological properties, density, and suspended particles.

It was found that at the lower projectile velocities, the high volume fraction SiO\textsubscript{2} suspensions resulted in the greatest change in velocity of the projectile, despite its lower density as compared to the other suspensions tested. This demonstrated the dominate effects of shear thickening on the dynamic behaviour at lower impact velocities. At high projectile velocities, fluid density became the dominant factor in the ballistic response of the suspensions.

\section*{Acknowledgements}
The authors would like to thank Jacques Blais of DRDC-Valcartier for his assistance in conducting the experimental work.

\section*{References}

\indent $^{1}$L.E. Gates Jr. (1968), AFML-TR-68-362.\\
$^{2}$Y.S. Lee, E.D. Wetzel, and N.J. Wagner (2003), J. Mat. Sci., \textbf{38}(13):2825-2833.\\
$^{3}$D.P. Kalman, R.L. Merrill, N.J. Wagner, and E.D. Wetzel (2009), Appl. Mat. \& Int. \textbf{1}(11):2602-2612.

\end{document}